\documentclass[twocolumn,pre,showpacs]{revtex4}

\usepackage{amssymb}
\usepackage{amsfonts}
\usepackage{amsmath}
\usepackage{graphicx}

\begin{document}
\title{Large deviations of ergodic counting processes: a statistical
mechanics approach}
\author{Adri\'{a}n A. Budini}
\affiliation{Consejo Nacional de Investigaciones Cient\'{\i}ficas y T\'{e}cnicas, Centro
At\'{o}mico Bariloche, Avenida E. Bustillo Km 9.5, (8400) Bariloche,
Argentina}
\date{\today}

\begin{abstract}
The large-deviation method allows to characterize an ergodic counting
process in terms of a thermodynamic frame where a free energy function
determines the asymptotic non-stationary statistical properties of its
fluctuations. Here, we study this formalism through a statistical mechanics
approach, i.e., with an auxiliary counting process that maximizes an entropy
function associated to the thermodynamic potential. We show that the
realizations of this auxiliary process can be obtained after applying a
conditional measurement scheme to the original ones, providing is this way
an alternative measurement interpretation of the thermodynamic approach.
General results are obtained for renewal counting processes, i.e., those
where the time intervals between consecutive events are independent and
defined by a unique waiting time distribution. The underlying statistical
mechanics is controlled by the same waiting time distribution, rescaled by
an exponential decay measured by the free energy function. A scale
invariance, shift closure, and intermittence phenomena are obtained and
interpreted in this context. Similar conclusions apply for non-renewal
processes when the memory between successive events is induced by a
stochastic waiting time distribution.
\end{abstract}

\maketitle

\section{Introduction}

In different scientific disciplines the measurement realizations of a given
process consist of a set of random points distributed along the real number
line. These \textquotedblleft stochastic point processes\textquotedblright\ 
\cite{isham,cox,daley,vanKampen} can be characterized through different
statistical objects and techniques. One of the most usual is to count the
(random) number of events up to a given point. For example, one can count up
to a given time the number of photons emitted by a quantum optical system 
\cite%
{MandelBook,zhengBloch,heSpectral,BarkaiExactMandel,slow,mukamel,budini,SMS,SMSJumps}%
, the number of electrons transported through a nanoscopic structure \cite%
{braggio,braggio1,braggio2}, or the number of spikes produced by a neuronal
system \cite{tuckwell,vreeswijk,lindner,nawrot,longtin}, just to name a few.
From these examples it becomes clear that the underlying (non-equilibrium)
dynamics that lead to the random processes may be of very different nature
and complexity.

A counting process can statistically be characterized by a set of functions,
each one defining the probability of n-events occurring up to a given time.
These objects completely characterize the fluctuations (moments and
cumulants) of the measurement realizations. The behavior (time-dependence)
of the counting probabilities is not universal and depends on each specific
case. Nevertheless, after a characteristic transient time that depends on
each system, a general characterization is expected to apply. In fact, when
the probabilities of a system scales as exponential functions (asymptotic
regime), the large deviation (LD) approach \cite{touchette} allows to read
the scaling rates in the context of an equilibrium thermodynamics formalism.
In this frame \cite%
{touchette,lecomte,vander,wijland,sollich,rolland,garrahan,Large}, the
entire (asymptotic) statistical properties of a process can be obtained from
a free energy function \cite{raquel}\ and characterized, in particular, from
its associated thermodynamic response functions (dynamical order
parameters), i.e., its derivatives with respect to an intensive parameter
(conjugate field). The thermodynamic frame describes in a unified way both
the typical fluctuations of a system, such as those that can be fit in a
Gaussian description [central limit theorem (CLT)], but also large deviation
fluctuations that go beyond the previous regime.

The LD theory and its associated thermodynamic frame were recently applied
for characterizing photon counting processes in different quantum optical
systems \cite{garrahan,Large}. The thermodynamic responses develop
properties such as scale invariant points, phase transitions and finite-size
effects, which in turn indicate drastic changes in the statistical
properties of the counting process.

While previous results \cite{garrahan,Large} confirm that the LD method
provides a deep description of a stochastic counting process, its associated
thermodynamic structure can only be analyzed after specifying a given
problem or situation. Furthermore, while the full thermodynamic formalism
depends on the conjugate (counting) field it is not completely clear how
this dependence may become observable or explicitly defined from the
statistics of the counting process. The main goal of this paper is to shed
light on this issue and to demonstrate that some general features of the
thermodynamic frame can be characterized without knowing the specific
properties and underlying dynamics (classical, quantum, phenomenological) of
the counting process. The present analysis applies to ergodic
(unidirectional) processes.

Instead of focusing on the thermodynamic properties, which depend on each
particular problem, here we analyze its associated statistical mechanics,
i.e., an auxiliary counting process that maximizes an entropy function
related to the free energy of the thermodynamic frame. We show that its
realizations can be obtained after applying a conditional measurement scheme
(Fig. 1) to the original ones. The role played by the conjugate field is
clarified in this context, providing an alternative measurement foundation
of the formalism and results developed in Refs. \cite{garrahan,Large}.
Moreover, some general results and conclusions follow from the measurement
scheme.

A general analysis is developed by assuming a renewal property \cite%
{isham,cox,daley,vanKampen}. Hence, the counting process is characterized by
a probability distribution, or waiting time distribution (WTD), that defines
the statistics of the random time intervals between consecutive events. On
the basis of the conditional measurement scheme, we show that the
thermodynamic frame can also be related to an underlying renewal counting
process whose WTD is the original one multiplied by an exponential decay
scaled by the free energy function. Furthermore, it is demonstrated that
scale invariant points, a shift closure property, intermittent and
(thermodynamic) finite-size effects can be developed under the renewal
hypothesis. While no general results can be formulated for non-renewal
counting process, we show that similar conclusions can be obtained for
process defined by a stochastic WTD. This non-renewal case arises in
different situations such as quantum optical systems \cite{SMSJumps} and
neuronal ones \cite{vreeswijk}.

The paper is outlined as follows. In Sec. II, after reviewing the LD
approach in the context of counting processes, we define and analyze the
underlying statistical mechanics associated to the thermodynamic formalism.
In Sec. III we study the central case of renewal counting process. In Sec.
IV we analyze non-renewal processes defined by a stochastic WTD. In Sec. V
we provide the conclusions.

\section{Large deviation theory of counting processes}

A counting process can be statistically characterized by a set of
probabilities $\{P_{n}(t)\}_{n=0}^{\infty },$ satisfying $0\leq P_{n}(t)\leq
1,$ and the normalization $\sum_{n=0}^{\infty }P_{n}(t)=1.$ Each $P_{n}(t)$
is the probability of occurrence of $n$-events up to time $t.$ As in Refs. 
\cite{Large,garrahan}, here we are restricting our analysis to
unidirectional counting processes, i.e., $n\geq 0.$

While the short time behavior of the counting probabilities depends on each
specific case, their asymptotic behavior may assume a universal structure.
For example, CLT allows to approximate their long time regime with a
Gaussian distribution \cite{isham,cox}%
\begin{equation}
\lim_{t\rightarrow \infty }P_{n}(t)\overset{clt}{\approx }\sqrt{\frac{1}{%
2\pi \overline{\Delta n_{t}^{2}}}}\exp \left\{ -\frac{[n-\bar{n}_{t}]^{2}}{2%
\overline{\Delta n_{t}^{2}}}\right\} ,  \label{gauss}
\end{equation}%
where $\bar{n}(t)$ is the average number 
\begin{equation}
\bar{n}_{t}=\sum_{n=0}^{\infty }nP_{n}(t)\approx \langle \langle n\rangle
\rangle t,
\end{equation}%
while $\overline{\Delta n_{t}^{2}}$ corresponds to the second cumulant 
\begin{equation}
\overline{\Delta n_{t}^{2}}=\sum_{n=0}^{\infty }(n-\bar{n}%
(t))^{2}P_{n}(t)\approx \langle \langle \Delta n^{2}\rangle \rangle t.
\end{equation}%
$\langle \langle n\rangle \rangle $ and $\langle \langle \Delta n^{2}\rangle
\rangle $\ are the asymptotic growing rates. For a fixed time, the
prediction of CLT is only valid up to a given accuracy in a central region
of the counting probabilities. By assuming valid an exponential asymptotic
behavior%
\begin{equation}
\lim_{t\rightarrow \infty }P_{n}(t)\approx \exp [-t\varphi (\frac{n}{t})],
\label{Pasimptotico}
\end{equation}%
LD method allows to describe the long time regime beyond the Gaussian
approximation. On the basis of a saddle-point approximation \cite{touchette}%
, the function $\varphi (n)$ can be obtained from a Legendre-Fenchel
transformation 
\begin{equation}
\varphi (n)=\max_{s}[\Theta (s)-sn],  \label{PhiN}
\end{equation}%
together with the inversion formula%
\begin{equation}
\Theta (s)=\min_{n}[\varphi (n)+sn].  \label{TetaS}
\end{equation}%
The function $\Theta (s)$ is defined by the asymptotic behavior%
\begin{equation}
\lim_{t\rightarrow \infty }Z(t,s)\approx \exp [-t\Theta (s)].
\label{Zasimptotico}
\end{equation}%
Here, $Z(t,s)$\ is the characteristic function associated to the counting
probabilities 
\begin{equation}
Z(t,s)\equiv \sum_{n=0}^{\infty }P_{n}(t)e^{-sn},  \label{partition}
\end{equation}%
where $s$ is a real dimensionless parameter. Hence, $\Theta (0)=0.$ From
relations (\ref{PhiN}) and (\ref{TetaS}) one deduces that any of the
functions $\varphi (n)$ or $\Theta (s)$ provide a complete description of
the asymptotic counting statistics.

CLT is covered by the LD approach. From Eq.~(\ref{gauss}) the LD rate
function $\varphi (n)$ can be approximated as%
\begin{equation}
\varphi (n)\overset{clt}{\approx }\frac{(n-\langle \langle n\rangle \rangle
)^{2}}{2\langle \langle \Delta n^{2}\rangle \rangle }.
\end{equation}%
Furthermore, Eq. (\ref{TetaS}) allows us to obtain%
\begin{equation}
\Theta (s)\overset{clt}{\approx }s\langle \langle n\rangle \rangle -\frac{%
s^{2}}{2}\langle \langle \Delta n^{2}\rangle \rangle .  \label{ThetaCLT}
\end{equation}%
This expression can be read as a Taylor expansion up to second order in $s$
of $\Theta (s),$ which in turn corresponds to a quadratic fitting of $%
\varphi (n)$ around its minimal value, $n=\langle \langle n\rangle \rangle .$
Thus, the Gaussian prediction of CLT follows from a quadratic approximation
of both $\varphi (n)$ and $\Theta (s).$ The LD method takes into account all
higher contributions related to higher cumulants.

\subsection{Underlying thermodynamics and statistical mechanics}

Legendre-Fenchel transformations (\ref{PhiN}) and (\ref{TetaS}) suggest to
analyze the LD method within an equilibrium thermodynamic frame \cite%
{garrahan,Large}. In fact, the functions $\varphi (n)$ and $\Theta (s)$ can
be read and related to the internal energy $(U)$ and grand potential $%
[\Theta (s)]$\ of a thermodynamic open system, which satisfy the relation 
\cite{raquel} 
\begin{equation}
\Theta (s)=U-TS+sN_{\#}.  \label{Grand}
\end{equation}%
$N_{\#}$ is the average \textquotedblleft particle
number,\textquotedblright\ $s$ plays the role of a (dimensionless) chemical
potential (intensive parameter),\ and the (dimensionless) temperature $T$
can be taken as one. Here, we re-derive these relations and associations 
\cite{Large} from an equivalent statistical mechanics approach. Hence, we
search for a set of auxiliary (counting) probabilities $\{q_{n}(t)\}_{n=0}^{%
\infty }$ that maximize an entropy function $S$ consistent with the
thermodynamic relation (\ref{Grand}).

As the LD approach applies in the asymptotic regime, we define a normalized
entropy function as 
\begin{equation}
S\equiv -\lim_{t\rightarrow \infty }\frac{1}{t}\sum_{n=0}^{\infty
}q_{n}(t)\log [q_{n}(t)].  \label{Entropy}
\end{equation}%
In order to find the probabilities $\{q_{n}(t)\}_{n=0}^{\infty },$ we
maximize the entropy function (at a fixed time) under the constraints of
probability normalization $\sum_{n=0}^{\infty }q_{n}(t)=1,$ fixed average
(particle) number 
\begin{equation}
N_{\#}=\left\langle \left\langle N\right\rangle \right\rangle \equiv
\lim_{t\rightarrow \infty }\frac{1}{t}\sum_{n=0}^{\infty }q_{n}(t)n,
\label{Naverage}
\end{equation}%
and (average) internal energy 
\begin{equation}
U=\left\langle \left\langle E\right\rangle \right\rangle \equiv
-\lim_{t\rightarrow \infty }\frac{1}{t}\sum_{n=0}^{\infty }q_{n}(t)\log
[P_{n}(t)].  \label{InternalEnergy}
\end{equation}%
By using the method of Lagrange multipliers \cite{raquel}, after defining
the free energy or grand potential [see Eq. (\ref{Zasimptotico})]%
\begin{equation}
\Theta (s)=-\lim_{t\rightarrow \infty }\frac{1}{t}\log [Z(t,s)],
\label{FreeEnergy}
\end{equation}%
and using the thermodynamic relation Eq. (\ref{Grand}), we get the
asymptotic behavior 
\begin{equation}
\lim_{t\rightarrow \infty }q_{n}(t)\approx \exp \{-t[(\frac{n}{t})s+\varphi (%
\frac{n}{t})-\Theta (s)]\}.  \label{qAsimptotico}
\end{equation}%
In this derivation, we assumed units of energy such that the Lagrange
multiplier associated to the internal energy constraint (inverse of the
temperature) takes the value one ($T\rightarrow 1$), and $s$ corresponds to
the Lagrange multiplier associated to the constraint (\ref{Naverage}), i.e.,
the dimensionless parameter $s$ plays the role of a chemical potential.

Eq. (\ref{qAsimptotico}) is satisfied if we define the probabilities $%
q_{n}(t,s)\equiv q_{n}(t)$ at any time as [see Eqs. (\ref{Pasimptotico}) and
(\ref{Zasimptotico})]%
\begin{equation}
q_{n}(t,s)\equiv \frac{1}{Z(t,s)}P_{n}(t)e^{-sn},  \label{q}
\end{equation}%
where the \textquotedblleft partition function\textquotedblright\ $Z(t,s)$
follows from Eq. (\ref{partition}). These probabilities define an extra
counting process, which is parametrized by the dimensionless chemical
potential $s.$ Due to this dependence, the set of stochastic realizations
consistent with $\{q_{n}(t,s)\}_{n=0}^{\infty }$ is named as the $s$%
-ensemble \cite{garrahan,Large}.

Relevant information about the fluctuations of the counting process defined
by the set $\{P_{n}(t)\}_{n=0}^{\infty }$\ is encoded in the thermodynamical
response functions \cite{raquel}, such as the first and second derivatives
of the grand potential with respect to $s$%
\begin{equation}
\langle \langle N\rangle \rangle =\frac{\partial }{\partial s}\Theta (s),\ \
\ \ \ \ \ \ \ \langle \langle \Delta N^{2}\rangle \rangle =-\frac{\partial
^{2}}{\partial s^{2}}\Theta (s).  \label{ResponseFunctions}
\end{equation}%
In fact, different (dynamical) phases, as well as thermodynamic transitions
between them, can be established by analyzing the dependence of these
objects with respect to the pseudo chemical potential \cite{garrahan,Large}.
Due to this reason, the first response function is alternatively denoted as
a \textquotedblleft \textit{dynamical order parameter},\textquotedblright\
while $s$ is quoted as its non-equilibrium \textquotedblleft \textit{%
conjugate (counting) field}\textquotedblright\ \cite{garrahan}.

Response functions (\ref{ResponseFunctions}) can be written as the
normalized [$\lim_{t\rightarrow \infty }(1/t)\cdots $] average value and
variance of the number of events associated to the set of probabilities $%
\{q_{n}(t,s)\}_{n=0}^{\infty }$ [see Eq. (\ref{Naverage})]. The same
property is valid for higher objects. Hence, the $s$-ensemble provides an
alternative basis for describing the thermodynamic frame. This fact defines
the present approach.

\subsection{Realizations of the s-ensemble: conditional counting scheme}

The point process associated to probabilities $\{P_{n}(t)\}_{n=0}^{\infty }$
is our observable system. Therefore, one has access to its ensemble of
realizations. Nevertheless, it is not clear which kind of point process
leads to the counting process defined by the probabilities $%
\{q_{n}(t,s)\}_{n=0}^{\infty },$ Eq. (\ref{q}). While its definition seems
to be rather abstract, here we provide a measurement scheme for generating
its ensemble of realizations from those of $\{P_{n}(t)\}_{n=0}^{\infty }.$
The dependence of the thermodynamic frame on the conjugate field $s$ is
clarified by this result, which in turn answers the main issue raised up in
the introduction.

When $s>0,$ we realize that $0<\exp [-s]<1.$ Therefore, this factor can be
read as a probability. In this case, the realizations associated to the
counting probabilities $\{q_{n}(t,s)\}_{n=0}^{\infty }$ in the interval $%
(0,t)$ can be obtained as follows. Each realization corresponding to $%
\{P_{n}(t)\}_{n=0}^{\infty }$ is split in two channels (upper and lower
arms, see Fig.~1) with probabilities $\exp [-s]$ and $(1-\exp [-s])$
respectively. If in the interval $(0,t)$ all events are selected in the
upper arm, that realization is taken as one of the set $\{q_{n}(t,s)%
\}_{n=0}^{\infty }$ [Fig. 1(a)]. If at least one event happens in the lower
arm, the realization is discarded [Fig. 1(b)]. Therefore, each realization
of $\{P_{n}(t)\}_{n=0}^{\infty }$ having $n-$events is selected as one of
the $\{q_{n}(t,s)\}_{n=0}^{\infty }$ with probability $\exp [-sn]$ and
discarded with probability $\sum_{k=1}^{n}\binom{n}{k}%
e^{-s(n-k)}(1-e^{-s})^{k}=(1-e^{-sn}).$ This conditional selection
corresponds the factor $\exp [-sn]$ in the definition (\ref{q}).
Furthermore, $Z(t,s)$ can be read as the fraction of the total realizations
of $\{P_{n}(t)\}_{n=0}^{\infty }$ that are taken as realizations of the set $%
\{q_{n}(t,s)\}_{n=0}^{\infty }.$ Therefore, it provides the right
normalization associated to the conditional selection. Alternatively, it can
be read as the probability of not occurring any event in the lower arm in
the interval $(0,t).$ 
\begin{figure}[tbp]
\includegraphics[bb=130 336 650 620,angle=0,width=7.5cm]{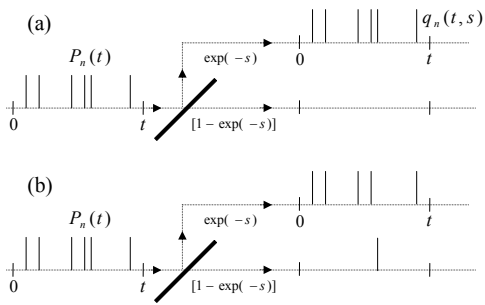} 
\caption{Conditional counting scheme that leads to the realizations of $%
\{q_{n}(t,s)\}_{n=0}^{\infty }.$ Each event of $\{P_{n}(t)\}_{n=0}^{\infty }$
(left vertical lines) is selected with probabilities $\exp [-s]$ and $%
(1-\exp [-s])$ in the upper and lower arms respectively. In (a) all events
went in the upper arm, leading to a s-realization. (b) Realizations with at
least one event in the lower arm are discarded.}
\end{figure}

When $s<0,$ it follows $\exp [-s]>1.$ Therefore, the previous scheme does
not apply. Nevertheless, a similar interpretation can be established. We
realize that for any value of $s$ it is possible to invert Eq. (\ref{q}) as%
\begin{equation}
P_{n}(t)=\frac{1}{Z_{q}(t,-s)}q_{n}(t,s)e^{sn},  \label{inversion}
\end{equation}%
where we have defined the $q-$partition function%
\begin{equation}
Z_{q}(t,s)\equiv \sum_{n=0}^{\infty }q_{n}(t,s)e^{-sn}.  \label{qPartition}
\end{equation}%
Notice that the normalization of the counting probabilities implies the
relation $Z_{q}(t,-s)=1/Z(t,s).$ For $s<0,$\ Eq. (\ref{inversion}) has the
same structure than Eq. (\ref{q}). Hence, even in this case, the previous
measurement scheme makes sense. Nevertheless, for $s<0$ the input signal
must correspond to the realizations of $\{q_{n}(t,s)\}_{n=0}^{\infty }$
while the the output signal delivers the realizations defined by the set $%
\{P_{n}(t)\}_{n=0}^{\infty }.$ Although this association is exact $(s<0),$
it does not provide a procedure for generating the $s-$ensemble from the
original counting process. In fact, it solves the inverse problem.

The relation between the original ensemble and the $s$-ensemble can be
extended to two arbitrary values, $s$ and $s^{\prime },$ of the pseudo
chemical potential. We write%
\begin{equation}
q_{n}(t,s)=\frac{1}{Z(t,s,s^{\prime })}q_{n}(t,s^{\prime })e^{-(s-s^{\prime
})n},
\end{equation}%
where the normalization here reads%
\begin{equation}
Z(t,s,s^{\prime })\equiv \sum_{n=0}^{\infty }q_{n}(t,s^{\prime
})e^{-(s-s^{\prime })n}.
\end{equation}%
For $s>s^{\prime },$ the realizations of $\{q_{n}(t,s)\}_{n=0}^{\infty }$
follow from those of $\{q_{n}(t,s^{\prime })\}_{n=0}^{\infty }$ by applying
the conditional measurement procedure sketched in Fig.~1. The set $%
\{P_{n}(t)\}_{n=0}^{\infty }$ is associated to $s^{\prime }=0.$ Furthermore,
it is simple to realize why it is not possible to build up a measurement
scheme for obtaining the realizations of $\{q_{n}(t,s^{\prime
})\}_{n=0}^{\infty }$ from those of $\{q_{n}(t,s)\}_{n=0}^{\infty },$ i.e.,
the case $s<s^{\prime }.$ In fact, the realizations of the last counting
process follow from a subset of the realizations of the former one, implying
a lack of information which cannot be recovered from any measurement scheme.

From the definition of the conditional measurement scheme (Fig. 1) and the
previous analysis, it is simple to realize that the thermodynamic response $%
\langle \langle N\rangle \rangle $ [Eq. (\ref{Naverage})] is a strictly
decreasing function of $s,$ implying%
\begin{equation}
\lim_{s\rightarrow +\infty }\langle \langle N\rangle \rangle =0,\ \ \ \ \ \
\ \ \lim_{s\rightarrow -\infty }\langle \langle N\rangle \rangle =\infty .
\label{ResponseLimits}
\end{equation}%
These constraints imply the concavity of the free energy function $\Theta
(s) $ and the limits%
\begin{equation}
\lim_{s\rightarrow +\infty }\Theta (s)=|C|,\ \ \ \ \ \ \ \
\lim_{s\rightarrow -\infty }\Theta (s)=-\infty ,  \label{limits}
\end{equation}%
where $C$ is a real constant. Consistently with the behavior of $\langle
\langle N\rangle \rangle ,$ in general $\langle \langle \Delta N^{2}\rangle
\rangle $ is neither an increasing nor a decreasing function of $s.$

The function $Z(t,s)$ measures the fraction of the number of realizations
from the original counting process that are taken for the $s$-ensemble. The
asymptotic behavior (\ref{Zasimptotico}) implies that a reliable measurement
of the set $\{q_{n}(t,s)\}_{n=0}^{\infty }$ [Eq. (\ref{q})] and the response
functions involves a number of realizations of $\{P_{n}(t,s)\}_{n=0}^{\infty
}$ that scales as $\exp [t^{\ast }\Theta (s)],$ where $t^{\ast }$ is larger
than the transitory time interval. From a numerical point of view, this
condition is very restrictive. In fact, it can take a long time before a
realization is generated. In spite of this severe limitation, below (Fig. 3)
we show that the measurement scheme can in fact be numerically implemented
for moderate values of $s,$ i.e., $s\approx 0,$ such that the product $%
t^{\ast }\Theta (s)$ does not assume very high values.

\section{Renewal counting processes}

While the thermodynamic potential $\Theta (s)$ provides a complete
description of LD fluctuations, its functional form depends on each specific
case. In contrast, some general results can be established from the
statistical mechanics formulation based on the set $\{q_{n}(t,s)\}_{n=0}^{%
\infty }.$ Here, we use this advantage for providing a general
characterization of an arbitrary renewal counting process \cite{isham,cox}.

A renewal point process is defined by a WTD satisfying $w(t)\geq 0$ and the
normalization $\int_{0}^{\infty }dt^{\prime }w(t^{\prime })=1.$ It defines
the probability distribution of the (random) time interval between
consecutive events. Therefore, the counting probabilities $%
\{P_{n}(t,s)\}_{n=0}^{\infty }$ can be written as \cite{godreche} 
\begin{subequations}
\label{RenewalCounting}
\begin{eqnarray}
P_{0}(t) &=&1-\int_{0}^{t}dt^{\prime }w(t^{\prime }), \\
P_{n}(t) &=&\int_{0}^{t}dt^{\prime }w(t-t^{\prime })P_{n-1}(t^{\prime }).
\end{eqnarray}%
By working in a Laplace domain, it is simple to derive the equivalent
evolution equations 
\end{subequations}
\begin{subequations}
\label{PnEvolution}
\begin{eqnarray}
\frac{d}{dt}P_{0}(t) &=&-\int_{0}^{t}dt^{\prime }K(t-t^{\prime
})P_{0}(t^{\prime }), \\
\frac{d}{dt}P_{n}(t) &=&-\int_{0}^{t}dt^{\prime }K(t-t^{\prime
})[P_{n}(t^{\prime })-P_{n-1}(t^{\prime })].\ \ \ \ \ \ 
\end{eqnarray}%
The memory kernel $K(t)$ is defined in the Laplace domain, $%
g(u)=\int_{0}^{\infty }dtg(t)\exp [-ut],$ as 
\end{subequations}
\begin{equation}
K(u)=\frac{w(u)}{P_{0}(u)}=\frac{uw(u)}{1-w(u)}.  \label{kernel}
\end{equation}%
While the kernel $K(t)$\ does not have a straightforward physical
interpretation, its time integral does. After defining $f(t)\equiv
\int_{0}^{t}dt^{\prime }K(t^{\prime }),$ or equivalently in the Laplace
domain $f(u)=K(u)/u,$ from Eq. (\ref{kernel}) we get%
\begin{equation}
f(u)=\frac{w(u)}{1-w(u)}=\sum_{n=1}^{\infty }[w(u)]^{n}.  \label{sprinkling}
\end{equation}%
By writing this expression in the time domain in terms of successive
convolutions of $w(t),$ we realize that $f(t)dt$ provides the probability of
observing an event in $(t,t+dt)$ independently of the occurrence of any
extra events in previous time interval $(0,t)$ \cite{godreche}. In contrast, 
$w(t)dt$ has the same interpretation under the condition that not any event
occurs in $(0,t).$

\subsection{s-ensemble}

As the renewal property implies the absence of memory between consecutive
events, the $s$-ensemble (\ref{q}) can easily be characterized. From Eq. (%
\ref{PnEvolution}), the evolution of the partition function (\ref{partition}%
) reads 
\begin{equation}
\frac{d}{dt}Z(t,s)=-(1-e^{-s})\int_{0}^{t}dt^{\prime }K(t-t^{\prime
})Z(t^{\prime },s),  \label{ZNonLocalEvolution}
\end{equation}%
whose solution in the Laplace domain is%
\begin{equation}
Z(u,s)=\frac{1}{u+K(u)(1-e^{-s})}.  \label{ZetalLaplace}
\end{equation}%
By writing the survival probability in the Laplace domain as $%
P_{0}(u)=[u+K(u)]^{-1}$ [see Eq. (\ref{PnEvolution})], it follows that $%
Z(u,s)$ with $s>0$ can be read as the Laplace transform of a survival
probability defined with the normalized kernel $K(u)\rightarrow
(1-e^{-s})K(u),$ which in turn implies $f(u)\rightarrow (1-e^{-s})f(u).$
Therefore, we confirm that in fact $Z(t,s)$ can be identified with the
survival probability of the lower arm of the measurement scheme of Fig.~1.

From its definition and using the renewal property, Eq.~(\ref%
{RenewalCounting}), $Z(t,s)$ can alternatively be written as 
\begin{equation}
Z(t,s)=P_{0}(t)+\int_{0}^{t}dt^{\prime }w(t-t^{\prime })e^{-s}Z(t^{\prime
},s).  \label{ZetaZeta}
\end{equation}%
Then, as $q_{0}(t,s)=P_{0}(t)/Z(t,s)$ it follows 
\begin{equation}
q_{0}(t,s)=1-\int_{0}^{t}dt^{\prime }w_{q}(t,t^{\prime },s).
\label{qoRenewal}
\end{equation}%
Furthermore, from Eqs. (\ref{q}) and (\ref{RenewalCounting}), for $n\geq 1$
we get%
\begin{equation}
q_{n}(t,s)=\int_{0}^{t}dt^{\prime }w_{q}(t,t^{\prime },s)q_{n-1}(t^{\prime
},s),  \label{qnRenewal}
\end{equation}%
where we have defined 
\begin{equation}
w_{q}(t,t^{\prime },s)\equiv e^{-s}w(t-t^{\prime })\frac{Z(t^{\prime },s)}{%
Z(t,s)}.  \label{qWaiting}
\end{equation}%
This expression, together with relations (\ref{qoRenewal}) and (\ref%
{qnRenewal}), imply that the $s$-ensemble is a non-stationary renewal
counting process [compare with Eqs. (\ref{RenewalCounting})]. In fact, here
the WTD $w_{q}(t,t^{\prime },s)$ not only depends on the interval between
successive events, but also on the time of the last event. Consistently with
the measurement scheme of Fig. 1, this dependence is introduced by the
partition function factors. By introducing the interval $\tau =t-t^{\prime }$
between two consecutive events, we write $w_{q}(t,t-\tau ,s)=e^{-s}w(\tau
)Z(t-\tau ,s)/Z(t,s).$ In this way, $w_{q}(t,t-\tau ,s)$ can be read as the
probability density for observing an event at $t$ given that the last one
occurred at time $(t-\tau ).$ The normalization $\lim_{t\rightarrow \infty
}\int_{0}^{t}d\tau w_{q}(t,t-\tau ,s)=1,$ follows from Eqs. (\ref{ZetaZeta})
and (\ref{qoRenewal}) after using that $\lim_{t\rightarrow \infty
}q_{0}(t,s)=0.$ In the long time regime, where the partition function can be
approximated by an exponential decay, Eq. (\ref{Zasimptotico}), we get%
\begin{eqnarray}
w_{q}^{\infty }(\tau ,s) &\equiv &\lim_{t\rightarrow \infty }w_{q}(t,t-\tau
,s),  \notag \\
&=&e^{-s}w(\tau )\exp [\tau \Theta (s)].  \label{waitingSEstacionaria}
\end{eqnarray}%
Therefore, in the regime where the LD approach applies, the counting process
defined by the set $\{q_{n}(t,s)\}_{n=0}^{\infty }$ becomes a stationary
renewal one, which in turn is controlled by the WTD $w_{q}^{\infty }(\tau
,s).$ This is one of the central results of this section. Notice that it is
valid for any value of $s.$ On the other hand, one can also associate to the 
$s$-ensemble a non-conditional probability distribution%
\begin{equation}
f_{q}(t,t^{\prime },s)\!=\!w_{q}(t,t^{\prime },s)+\!\int_{t^{\prime
}}^{t}\!\!dt_{1}w_{q}(t,t_{1},s)f_{q}(t_{1},t^{\prime },s),  \label{fq}
\end{equation}%
which defines the probability density of observing an event in $(t,t+dt)$
given that one occurred at $t^{\prime }$ and independently of the occurrence
of any extra events in the time interval $(t,t^{\prime }).$ By solving Eq. (%
\ref{fq}) iteratively, after some calculations we get%
\begin{equation}
f_{q}(t,t^{\prime },s)=e^{-s}\digamma (t-t^{\prime },s)\frac{Z(t^{\prime },s)%
}{Z(t,s)},
\end{equation}%
where $\digamma (t,s)$ is defined in the Laplace domain as%
\begin{equation}
\digamma (u,s)=K(u)Z(u,s)=\frac{K(u)}{u+K(u)(1-e^{-s})}.
\end{equation}%
Consistently, $f_{q}(t,t^{\prime },0)=f(t-t^{\prime }).$ Furthermore, in
correspondence with Eq. (\ref{waitingSEstacionaria}), we define $%
f_{q}^{\infty }(\tau ,s)=\lim_{t\rightarrow \infty }f_{q}(t,t-\tau ,s),$
which reads%
\begin{equation}
f_{q}^{\infty }(\tau ,s)=e^{-s}\digamma (\tau ,s)\exp [\tau \Theta (s)].
\label{efeEstacionaria}
\end{equation}%
From its definition \cite{godreche}, it follows the relation $\lim_{\tau
\rightarrow \infty }f_{q}^{\infty }(\tau ,s)=\langle \langle N\rangle
\rangle ,$ Eq. (\ref{Naverage}).

\subsection{Free energy and response functions}

The underlying renewal property of the $s$-ensemble allows us to get some
general expressions for the free energy and its response functions. The
Laplace transform of $w_{q}^{\infty }(\tau ,s)$ [Eq. (\ref%
{waitingSEstacionaria})] with respect to $\tau $ can be written as $%
w_{q}^{\infty }(u,s)=e^{-s}w(u-\Theta (s)).$ As this function satisfies the
normalization $\int_{0}^{\infty }d\tau w_{q}^{\infty }(\tau ,s)=1,$ or
equivalently $w_{q}^{\infty }(u,s)|_{u=0}=1,$ we obtain%
\begin{equation}
w(u-\Theta (s))|_{u=0}=e^{s}.  \label{WaitTeta}
\end{equation}%
By using this relation and the asymptotic values (\ref{limits}) one can
deduce the relation%
\begin{equation}
\lim_{s\rightarrow -\infty }\Theta (s)\approx \gamma _{0}(1-e^{-\alpha
_{0}s}),  \label{LimitTheta}
\end{equation}%
valid when $w(t)$ has a characteristic short time scale, i.e., $%
\lim_{u\rightarrow \infty }w(u)\simeq (\gamma _{0}/u)^{1/\alpha _{0}}.$


The Laplace transform of $w(t)$ also provides an equation for getting the
free energy function. In fact, it is the smaller solution of Eq. (\ref%
{WaitTeta}). This relation is valid for any (ergodic) renewal process.
Notice that the same kind of relations follow from definition (\ref%
{Zasimptotico}). In fact, by using the residues theorem, $\Theta (s)$ can be
determined as the smaller root of the denominator of Eq. (\ref{ZetalLaplace}%
).

As the process defined by the set $\{q_{n}(t,s)\}_{n=0}^{\infty }$ is a
renewal one in the stationary regime, the thermodynamic response functions
[Eq. (\ref{ResponseFunctions})] can be derived from $w_{q}^{\infty }(\tau
,s).$ In fact, by defining the time interval moments%
\begin{equation}
\tau _{n}(s)\equiv \int_{0}^{\infty }w_{q}^{\infty }(\tau ,s)\tau ^{n}d\tau ,
\end{equation}%
and by using the renewal property, it follows%
\begin{equation}
\langle \langle N\rangle \rangle =\frac{1}{\tau _{1}(s)},\ \ \ \ \ \ \ \ \ \
\ \ \langle \langle \Delta N^{2}\rangle \rangle =\frac{\tau _{2}(s)-[\tau
_{1}(s)]^{2}}{[\tau _{1}(s)]^{3}},  \label{ResponseRenewal}
\end{equation}%
where we have used that $\langle \langle N\rangle \rangle $ and $\langle
\langle \Delta N^{2}\rangle \rangle $ measure the linear growth of the first
two cumulants of the counting process defined by the set $%
\{q_{n}(t,s)\}_{n=0}^{\infty }$ (see page 496 of Ref. \cite{godreche}).
Similar relations are valid for higher response functions, i.e., $%
w_{q}^{\infty }(\tau ,s)$ completely determine the thermodynamic formalism.

\subsection{Applications}

Below, we show some applications that rely on the previously developed
results.

\subsubsection{Scale invariant renewal process}

In Ref. \cite{garrahan,Large}, a scale invariant property was found in a
class of photon counting process. It is defined by the condition%
\begin{equation}
\frac{\langle \langle \Delta N^{2}\rangle \rangle }{\langle \langle N\rangle
\rangle }=\alpha ,  \label{ScaleInvariant}
\end{equation}%
where the response functions follow from Eq. (\ref{ResponseFunctions}), and $%
\alpha $ is a positive real constant, $0<\alpha <\infty .$ Hence, the
normalized fluctuations do not depend on $s.$ The counting processes $%
\{P_{n}(t)\}_{n=0}^{\infty }$ and $\{q_{n}(t,s)\}_{n=0}^{\infty },$ for any
value of $s,$ are then sub-Poissonian for $0<\alpha <1,$ while for $1<\alpha
<\infty ,$ becomes super-Poissonian \cite{isham,cox,daley}. Here, we
demonstrate that the scale invariant property is always satisfied by a class
of (gamma) renewal process \cite{isham,cox,daley}, which are defined by the
WTD%
\begin{equation}
w(u)=\left( \frac{\gamma }{u+\gamma }\right) ^{1/\alpha }.
\label{WaiterInvariateEscalaLaplace}
\end{equation}%
In the time domain it reads%
\begin{equation}
w(t)=\frac{1}{\Gamma (1/\alpha )}\left( \gamma t\right) ^{\frac{1}{\alpha }%
-1}\gamma e^{-\gamma t}.  \label{WaiterInvariante}
\end{equation}%
The average interval between consecutive events is given by $%
\int_{0}^{\infty }w(t)tdt=(\alpha \gamma )^{-1}.$ By working the expressions
(\ref{RenewalCounting}) in the Laplace domain, the counting probabilities
read%
\begin{equation}
P_{n}(t)=\frac{\Gamma (\frac{n+1}{\alpha },\gamma t)}{\Gamma (\frac{n+1}{%
\alpha })}-\frac{\Gamma \lbrack \frac{n}{\alpha },\gamma t]}{\Gamma (\frac{n%
}{\alpha })}(1-\delta _{n0}),  \label{GammaDistribution}
\end{equation}%
where $\Gamma \lbrack k,x]$ is the incomplete gamma function $\Gamma \lbrack
k,x]\equiv \int_{x}^{\infty }z^{k-1}e^{-z}dz,$ and $\Gamma (k)$ is the Euler
gamma function, i.e., $\Gamma (k)=\Gamma \lbrack k,0].$ In the long time
regime, $\gamma t\gg 1,$ it is valid the approximation%
\begin{equation}
P_{n}(t)\simeq \frac{1}{\alpha \Gamma (\frac{n+1}{\alpha })}(\gamma t)^{%
\frac{n+1}{\alpha }-1}\exp [-\gamma t].
\end{equation}%
From this expression, it follows the LD function [Eq. (\ref{Pasimptotico})]%
\begin{equation}
\varphi (n)=\gamma \Big{\{}1-\frac{n}{\alpha \gamma }\Big{[}1-\log \Big{(}%
\frac{n}{\alpha \gamma }\Big{)}\Big{]}\Big{\}},
\end{equation}%
$(n/t)\rightarrow n.$ Its Legendre-Fenchel transformation [Eq. (\ref{TetaS}%
)] leads to the grand potential%
\begin{equation}
\Theta (s)=\gamma (1-e^{-\alpha s}).  \label{freeEnergy}
\end{equation}%
Consistently, this result straightforwardly follows from Eq. (\ref{WaitTeta}%
). Alternatively, Eq. (\ref{ZetalLaplace}) leads to the partition function%
\begin{equation}
Z(u,s)=\frac{1}{u}\frac{(u+\gamma )^{1/\alpha }-\gamma ^{1/\alpha }}{%
(u+\gamma )^{1/\alpha }-\gamma ^{1/\alpha }e^{-s}}.
\end{equation}%
From the residues theorem, the free energy (\ref{freeEnergy}) is recovered.
In fact, $u=-\Theta (s)$ cancels the denominator of this expression. The
grand potential (\ref{freeEnergy}) implies the thermodynamic response
functions [Eq. (\ref{ResponseFunctions})]%
\begin{equation}
\langle \langle N\rangle \rangle =\alpha \gamma e^{-\alpha s},\ \ \ \ \ \ \
\ \ \langle \langle \Delta N^{2}\rangle \rangle =\alpha ^{2}\gamma
e^{-\alpha s}.
\end{equation}%
Trivially, these expressions satisfy the invariant property Eq. (\ref%
{ScaleInvariant}).

While it is not possible to find a simple expression for the non-stationary
WTD $w_{q}(t,t^{\prime },s)$ [Eq. (\ref{qWaiting})], the stationary one [Eq.
(\ref{waitingSEstacionaria})]\ reads%
\begin{equation}
w_{q}^{\infty }(\tau ,s)=\frac{1}{\Gamma (1/\alpha )}\left( \gamma _{s}\tau
\right) ^{\frac{1}{\alpha }-1}\gamma _{s}\exp [-\gamma _{s}\tau ],
\end{equation}%
where the renormalized rate reads%
\begin{equation}
\gamma _{s}\equiv \gamma \exp [-\alpha s].  \label{RateCorrida}
\end{equation}%
The stationary WTD $w_{q}^{\infty }(\tau ,s)$ is the same as the original
one [see Eq. (\ref{WaiterInvariante})] under the replacement $\gamma
\rightarrow \gamma e^{-\alpha s}.$ Thus, the counting process defined by the
set $\{q_{n}(t,s)\}_{n=0}^{\infty },$ in the asymptotic regime, can be
obtained from the original one after a time rescaling%
\begin{equation}
\lim_{t\rightarrow \infty }q_{n}(t,s)\simeq \lim_{t\rightarrow \infty
}P_{n}(te^{-\alpha s}).  \label{ShiftTime}
\end{equation}%
This condition was derived from the WTD Eq. (\ref%
{WaiterInvariateEscalaLaplace}). Nevertheless, it is simple to prove that
its fulfilment is sufficient to guaranty the invariance property (\ref%
{ScaleInvariant}). By using that $Z_{q}(t,-s)=1/Z(t,s)$ [see Eq. (\ref%
{qPartition})], from (\ref{ShiftTime}) it follows the equivalent condition 
\begin{equation}
\Theta (-s)=-\Theta (s)e^{\alpha s},  \label{constraint}
\end{equation}%
which in turn allows to write an equation for $(d/ds)\Theta (s).$ Under the
constraints (\ref{constraint}), continuous derivatives in $s=0,$ and $\Theta
(0)=0,$ it has a unique non-null solution that is given by Eq. (\ref%
{freeEnergy}). Notice that this derivation does not rely on the renewal
property. Nevertheless, the previous one guarantees that $%
\{P_{n}(t)\}_{n=0}^{\infty }$ asymptotically converges to a renewal process
defined by the WTD (\ref{WaiterInvariateEscalaLaplace}).

When $\alpha =1,$ the WTD (\ref{WaiterInvariateEscalaLaplace}) becomes an
exponential one, $w(t)=\gamma \exp (-\gamma t).$ Thus, $f(t)=\gamma \theta
(t)$ [Eq. (\ref{sprinkling})] where $\theta (t)$ is the step function,
implying a local in time evolution $[K(t)=\gamma \delta (t)]$ of the
counting probabilities [Eq. (\ref{PnEvolution})]. Their solution read $%
P_{n}(t)=(\gamma t)^{n}\exp [-\gamma t]/n!,$ i.e., the very well-known
Poisson process \cite{isham,cox,daley,vanKampen}. The partition function
[Eq. (\ref{partition})] reads%
\begin{equation}
Z(t,s)=\exp [-\gamma t(1-e^{-s})],\ \ \ \ \alpha =1,  \label{ZPoisson}
\end{equation}%
which in turn implies the free energy (\ref{freeEnergy}) and its associated
response functions with $\alpha =1.$ The $s-$ensemble [Eq. (\ref{q})] at
\textquotedblleft any particular time\textquotedblright\ can easily be
written as $q_{n}(t,s)=(e^{-s}\gamma t)^{n}\exp [-e^{-s}\gamma t]/n!.$
Consistently, these probabilities also correspond to a Poisson process with
the normalized rate $\gamma \rightarrow e^{-s}\gamma .$ Its associated
non-stationary WTD [Eq. (\ref{qWaiting})] is stationary at all times%
\begin{equation}
w_{q}(t,t^{\prime },s)=(\gamma e^{-s})\exp [-(\gamma e^{-s})(t-t^{\prime
})],\ \ \ \ \alpha =1,
\end{equation}%
i.e., at any particular time it only depends on the difference $(t-t^{\prime
}).$ This property is only valid for the Poisson process, $\alpha =1.$

\textit{Scale invariant photon-emission process. }In Ref. \cite{garrahan} it
was analyzed the case of a two-level fluorescent system, where an external
laser field induces a continuous emission of photons at random times. The
photon counting process is a renewal one, being defined by the WTD%
\begin{equation}
w(u)=\frac{\gamma /2}{u+\gamma /2}\frac{\Omega ^{2}}{u^{2}+u\gamma +\Omega
^{2}}.  \label{WaitDet}
\end{equation}%
Here, $u$ is the Laplace variable, $\gamma $ denotes the natural decay rate
of the system while the Rabi frequency $\Omega $ measures the system-laser
coupling. The average time is $\int_{0}^{\infty }w(t)tdt=[\gamma \Omega
^{2}/(\gamma ^{2}+2\Omega ^{2})]^{-1}.$ When $2\Omega =\gamma ,$ it was
found the invariant scale property (\ref{ScaleInvariant}) with $\langle
\langle \Delta N^{2}\rangle \rangle /\langle \langle N\rangle \rangle =1/3.$
In this situation, (\ref{WaitDet}) becomes%
\begin{equation}
w(u)=\left( \frac{\Omega }{u+\Omega }\right) ^{3},
\end{equation}%
which in the time domain leads to $w(t)=\frac{1}{2}\Omega ^{3}t^{2}\exp
[-\Omega t].$ These expressions and the previous results confirm that in
fact $\alpha =1/3$ [see Eq. (\ref{WaiterInvariateEscalaLaplace})], showing
the consistence of the present approach.

\subsubsection{Shift closure property}

With $\{\xi \}$ we denote the characteristic parameters (rates, etc.) that
determine a given counting process $\{P_{n}(t)\}_{n=0}^{\infty }.$
Evidently, its associated free energy depends on $\{\xi \},$ property
denoted as $[\Theta (s)]_{\{\xi \}}.$ We define a shift closure property by
the condition%
\begin{equation}
\lbrack \Theta (s)]_{\{\xi (s_{0})\}}=[\Theta (s+s_{0})-\Theta
(s_{0})]_{\{\xi \}},  \label{Closure}
\end{equation}%
where $\{\xi (s_{0})\}$ are new characteristic parameters that depend on $%
s_{0}.$ Note that $[\Theta (s+s_{0})-\Theta (s_{0})]_{\{\xi \}}$ is the free
energy function of the counting process $\{q_{n}(t,s_{0})\}_{n=0}^{\infty }.$
In fact, all response functions, i.e., its derivatives with respect to $s$
[Eq. (\ref{ResponseFunctions})], are the same as those of $\Theta (s)$
shifted by $-s_{0}.$ Hence, the condition (\ref{Closure}) tells us that $%
\{q_{n}(t,s_{0})\}_{n=0}^{\infty }$ (the $s_{0}$-ensemble) belongs to the
family of processes obtained from $\{P_{n}(t)\}_{n=0}^{\infty }$ by a change
of characteristic parameters, $\{\xi \}\rightarrow \{\xi (s_{0})\}.$
Consequently, the fulfilment of Eq. (\ref{Closure})\ allows us to generate
the realizations of the $s$-ensemble by changing the parameters of the
original counting process. Equivalently, it implies that the change of
parameters $\{\xi \}\rightarrow \{\xi (s_{0})\}$ in the definition of $%
\{P_{n}(t)\}_{n=0}^{\infty }$ only produce a shifting of all of its response
functions.

Scale invariant processes, Eq. (\ref{freeEnergy}), satisfy the shift closure
condition\ with the mapping $\{\gamma \}\rightarrow \{\gamma e^{-\alpha
s_{0}}\}$ [Eq. (\ref{RateCorrida})]. In general, it is very difficult to
check the validity of (\ref{Closure}). Nevertheless, for renewal processes
it can alternatively be written in terms of the stationary WTD (\ref%
{waitingSEstacionaria}) as%
\begin{equation}
w_{q}^{\infty }(\tau ,s)=[w(\tau )]_{\{\xi (s)\}}.  \label{WaiterShift}
\end{equation}%
Hence, $w_{q}^{\infty }(\tau ,s)$ follows from $w(\tau )$ after replacing $%
\{\xi \}\rightarrow \{\xi (s)\}.$ When satisfied, this equation is not only
equivalent to (\ref{Closure}) but also provides a simple way for obtaining
the parameters mapping $\{\xi \}\rightarrow \{\xi (s=s_{0})\}.$

A broad class of WTDs satisfy the shift closure property. For example,
consider in the Laplace domain $w(u)=\prod_{i}\lambda _{i}/(u+\lambda _{i}).$
By using the normalization condition (\ref{WaitTeta}), we get%
\begin{equation}
w_{q}^{\infty }(u,s)=\prod_{i}\frac{\lambda _{i}(s)}{u+\lambda _{i}(s)},\ \
\ \ \ \ \lambda _{i}(s)=\lambda _{i}-\Theta (s).  \label{fluoreto}
\end{equation}%
Notice that $w_{q}^{\infty }(u,s)$ have the same structure than $w(u)$ with
the rescaled parameters $\{\xi (s)\}=\{\lambda _{i}(s)\}.$ This simple
result guarantees the validity of Eq. (\ref{Closure}). In general, $\lambda
_{i}$ may be complex numbers $(Re[\lambda _{i}]>0),$ which in turn are
functions of experimental parameters. Eq. (\ref{WaitDet}) has this
structure, with $\lambda _{i}=\lambda _{i}(\gamma ,\Omega ^{2}).$ In this
case, we checked that $w_{q}^{\infty }(u,s)$ can be written as $w(u)$ after
replacing $\gamma \rightarrow \gamma (s)$ and $\Omega ^{2}\rightarrow \Omega
^{2}(s),$ where $\gamma (s)=\gamma -2\Theta (s),$ and $\Omega ^{2}(s)=\Omega
^{2}+\Theta (s)[\Theta (s)-\gamma ],$ where $\Theta (s)$ is defined by Eq.
(35) of Ref. \cite{Large}. Notice that the parameter transformation
satisfies $\gamma ^{2}(s)-4\Omega ^{2}(s)=\gamma ^{2}-4\Omega ^{2}.$

Another class of WTDs that satisfy the shifting condition (\ref{WaiterShift}%
) is given by $w(t)=\sum_{k}p_{k}\gamma _{k}e^{-\gamma _{k}t},$ with $%
\sum_{k}p_{k}=1,$ $0<p_{k}<1.$ Thus, in this case $\{\xi \}=\{\gamma
_{k},p_{k}\}.$ From Eq. (\ref{waitingSEstacionaria}), we get%
\begin{equation}
w_{q}^{\infty }(\tau ,s)=\sum_{k}p_{k}(s)\gamma _{k}(s)e^{-\gamma
_{k}(s)\tau },  \label{variasRates}
\end{equation}%
where the parameters $\{\xi (s)\}=\{\gamma _{k}(s),p_{k}(s)\}$ are%
\begin{equation}
\gamma _{k}(s)=\gamma _{k}-\Theta (s),\ \ \ \ \ \ p_{k}(s)=p_{k}\frac{%
e^{-s}\gamma _{k}}{\gamma _{k}-\Theta (s)},  \label{mapeoExpor}
\end{equation}%
with $\gamma _{k}(s)>0$ and $\sum_{k}p_{k}(s)=1.$ Notice that $w_{q}^{\infty
}(\tau ,s)$ can also be written as a superposition of Poisson distributions,
which guarantees the fulfilment of (\ref{Closure}) with the mapping $%
\{\gamma _{k},p_{k}\}\rightarrow \{\gamma _{k}(s_{0}),p_{k}(s_{0})\},$ Eq. (%
\ref{mapeoExpor}).

In order to exemplify the previous results, we consider a bi-exponential WTD 
\begin{equation}
w(t)=(1-p)\gamma _{>}e^{-\gamma _{>}t}+p\gamma _{<}e^{-\gamma _{<}t},
\label{WaitBiExpo}
\end{equation}%
where $\gamma _{>}$ and $\gamma _{<}$ are the characteristic rates and the
parameter $p$ satisfies $0\leq p\leq 1.$ The kernel (\ref{kernel}) defines
the evolution of the counting probabilities $\{P_{n}(t)\}_{n=0}^{\infty }$
[Eq. (\ref{PnEvolution})]. It reads%
\begin{equation}
K(t)=\gamma \{\delta (t)-\beta \exp [-\eta t]\},  \label{KernelBiExpo}
\end{equation}%
where the characteristic parameters are 
\begin{subequations}
\label{definitions}
\begin{eqnarray}
\gamma  &\equiv &(1-p)\gamma _{>}+p\gamma _{<}, \\
\beta  &\equiv &\{(1-p)\gamma _{>}^{2}+p\gamma _{<}^{2}-\gamma ^{2}\}/\gamma
, \\
\eta  &\equiv &(1-p)\gamma _{<}+p\gamma _{>}.
\end{eqnarray}%
The grand potential [Eq. (\ref{FreeEnergy})] follows from the relation (\ref%
{WaitTeta}). Equivalently, by using the residues theorem it can be obtained
from the smaller root of the denominator of Eq. (\ref{ZetalLaplace}), $%
u^{2}+u[\gamma (1-e^{-s})+\eta ]+(\eta -\beta )\gamma (1-e^{-s})=0.$ We get
\end{subequations}
\begin{eqnarray}
\Theta (s) &=&\frac{\gamma }{2}(1-e^{-s})+\frac{\eta }{2}-\frac{1}{2}\Big{\{}%
\lbrack \eta -\gamma (1-e^{-s})]^{2}  \notag \\
&&+4\beta \gamma (1-e^{-s})\Big{\}}^{1/2}.  \label{GrandBiExpo}
\end{eqnarray}%
\begin{figure}[tbp]
\includegraphics[bb=10 310 410 635,angle=0,width=7.5cm]{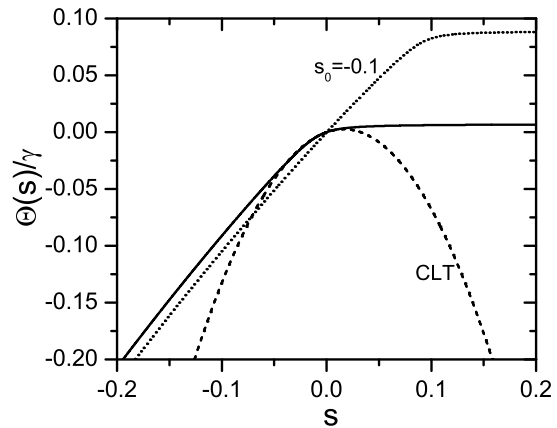} 
\caption{Free energy function (\protect\ref{GrandBiExpo}) (full line) in
terms of $s.$ The parameters of the WTD [Eq. (\protect\ref{WaitBiExpo})] are 
$\protect\gamma _{<}/\protect\gamma _{>}=0.007$ and $p=0.015$ $[\protect%
\gamma /\protect\gamma _{>}=0.9815].$ The dotted line corresponds to Eq. (%
\protect\ref{Closure}) with $s_{0}=-0.1.$ The short-dashed line is the
estimation based on the CLT, Eq. (\protect\ref{ThetaCLT}). }
\end{figure}
In Fig. 2 we plot $\Theta (s)$ (full line) as a function of the chemical
potential $s$ for a particular set of parameter values of the WTD (\ref%
{WaitBiExpo}). The limits (\ref{limits}) are satisfied. The dotted line
corresponds to $[\Theta (s+s_{0})-\Theta (s_{0})],$ with $s_{0}=-0.1.$
Consistently, we checked that the same (shifted) free energy function
follows from Eq. (\ref{GrandBiExpo}) under parameter replacements $\{\xi
\}=\{p,\gamma _{>},\gamma _{<}\}\rightarrow \{p(s_{0}),\gamma
_{>}(s_{0}),\gamma _{<}(s_{0})\}=\{\xi (s_{0})\},$ defined by Eq. (\ref%
{mapeoExpor}). This example explicitly shows the meaning of the shift
closure condition.

In Fig. 3 we plot the response functions, Eq. (\ref{ResponseFunctions}).
Consistently, they fulfill Eq. (\ref{ResponseRenewal}). We also checked that
the shifted response functions coincide with the derivative with respect to $%
s$ of $[\Theta (s)]_{\{\xi (s_{0})\}}.$ In Fig. 3(a) all plotted functions
are normalized by the rate $\gamma ,$ Eq. (\ref{definitions}), which is
different for each set of parameter values, $s=0,-0.1,-0.2.$ Hence, the
shifted response functions differ from the original one $(s=0)$\ in a
multiplicative factor. This effect is absent in the normalized second
response function $\langle \langle \Delta N^{2}\rangle \rangle /\langle
\langle N\rangle \rangle ,$ Fig. 3(b). 
\begin{figure}[tbp]
\includegraphics[bb=20 5 430 605,angle=0,width=7.5cm]{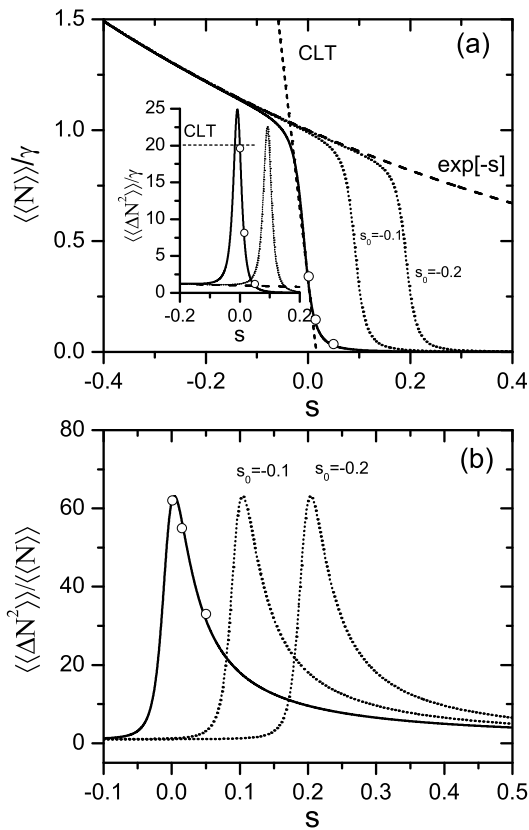} 
\caption{(a) First and second (inset) response functions, Eq. (\protect\ref%
{ResponseFunctions}). The parameters of the WTD are the same than in Fig. 2.
The dotted lines correspond to a set of different parameters values obtained
from Eq. (\protect\ref{mapeoExpor}) with $s_{0}=-0.1$ and $s_{0}=-0.2.$ The
circles correspond to the numerical values obtained from the measurement
scheme of Fig. 1 with the transformations $s=0\rightarrow s=0.001,$ $0.015,$ 
$0.05.$ The short-dashed lines are the estimations based on the CLT, Eq. (%
\protect\ref{ThetaCLT}). The long-dashed lines correspond to the response
functions of a Poisson process with rate $\protect\gamma $ (see text). (b)
Normalized second response function $\langle \langle \Delta N^{2}\rangle
\rangle /\left\langle \left\langle N\right\rangle \right\rangle .$}
\end{figure}

\subsubsection{Intermittent renewal process and finite-size effects}

In order to enlighten the LD method, in Figs. 2 and 3 we show the
predictions corresponding to the CLT (short-dashed lines), i.e., $\Theta (s)%
\overset{clt}{=}s(\langle \langle N\rangle \rangle
|_{s=0})-(s^{2}/2)(\langle \langle \Delta N^{2}\rangle \rangle |_{s=0}),$
Eq. (\ref{ThetaCLT}). Thus, the first response function has a linear
dependence on $s$ and the second one becomes a constant. While the first
approximation is valid in a small region around $s\approx 0,$ the second one
only applies at $s=0.$ 
\begin{figure}[tb]
\includegraphics[bb=65 0 500 620,angle=0,width=7.5cm]{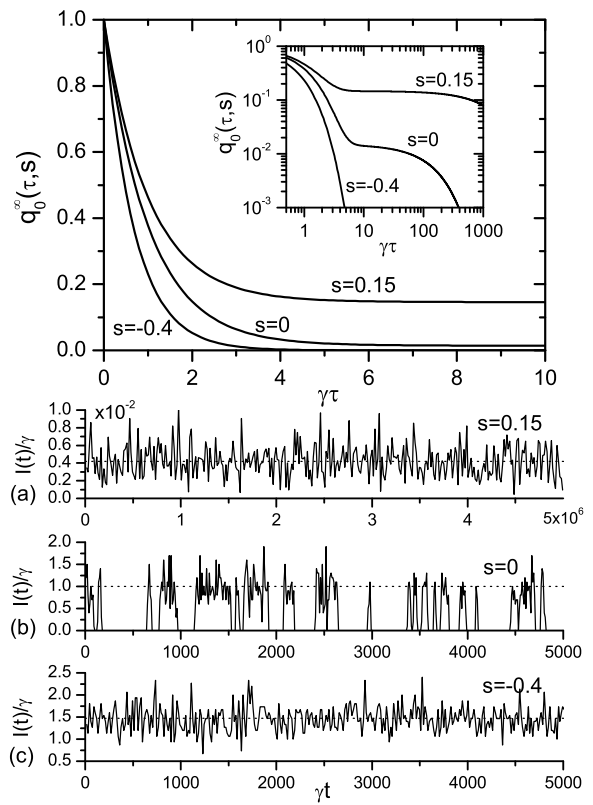}
\caption{Stationary survival probability (\protect\ref{qSurvival}) as a
function of time for different values of $s.$ The inset corresponds to a
log-log plot. The intensity realizations (see text) ploted in (a), (b), and
(c) correspond to $s=+0.15,$ $s=0.,$ and $s=-0.4$ respectively. The
parameters of the WTD (\protect\ref{WaitBiExpo}) are the same than in Fig.
2. }
\end{figure}

The previous analysis implies that the CLT approximation could miss a phase
transition within the thermodynamic approach. In contrast, the response
functions plotted in Fig. 3 develops properties consistent with finite-size
effects in a first order transition \cite{binderBook,landauIsing}. In fact,
the first response function changes abruptly around $s\approx 0,$ while the
second one develops a narrow peak. As shown in Refs. \cite{garrahan,Large}
these features can be related to an intermittence (blinking) property of the
counting process. Here, we show that renewal processes, independently of the
underlying dynamics, may also develop similar phenomena.

The main features of the response functions shown in Fig. 3 can be easily
interpreted in terms of the $s$-ensemble dynamics, which in turn is defined
by the stationary WTD (\ref{waitingSEstacionaria}). In Fig. 4, in order to
show the $s$-dependence of $w_{q}^{\infty }(\tau ,s),$ we plot its
associated survival probability 
\begin{equation}
q_{0}^{\infty }(\tau ,s)\equiv 1-\int_{0}^{\tau }d\tau ^{\prime
}w_{q}^{\infty }(\tau ^{\prime },s),  \label{qSurvival}
\end{equation}%
or equivalently $w_{q}^{\infty }(\tau ,s)=-(d/d\tau )q_{0}^{\infty }(\tau
,s).$ This function allows to generate the ensemble of realizations
consistent with the asymptotic behavior of the counting probabilities $%
\{q_{n}(t,s)\}_{n=0}^{\infty }.$ In fact, the time interval between
consecutive events can be determined by solving the equation $q_{0}^{\infty
}(\tau ,s)=r,$ where $r$ is a random number in the interval $(0,1).$

The previous algorithm allows us to check the measurement scheme defined in
Fig. 1. We numerically implemented the conditional selection to the
generated realizations $(s=0)$ and calculated numerically $\left\langle
\left\langle N\right\rangle \right\rangle $\ from the asymptotic behavior (%
\ref{Naverage}). $\langle \langle \Delta N^{2}\rangle \rangle $ can be
obtained in a similar way as the growing rate of the quadratic fluctuations.
In Fig. 3 we show the results (circles) for the transformations $%
s=0\rightarrow s=0.001,$ $0.015,$ $0.05.$ Even for values of $s$ where the
predictions of the CLT do not apply, the numerical estimations agree with
the theoretical results.

In Fig. 4, for each chosen value of $s,$ we also plotted the intensity $I(t)$
associated to an arbitrary stochastic realization of the $s$-ensemble. They
were obtained numerically from $q_{0}^{\infty }(\tau ,s).$ The intensity is
defined as $I(t)=(d/dt)n_{st}^{(s)}(t)\approx \lbrack n_{st}^{(s)}(t+\Delta
)-n_{st}^{(s)}(t)]/\Delta ,$ where $n_{st}^{(s)}(t)$ is the stochastic
number of events up to time $t$ and $\Delta $ define an adequate
discretization time. Consistently, one can always find a $\Delta $ where $%
I(t)$ fluctuates around $\left\langle \left\langle N\right\rangle
\right\rangle $ (Fig. 4(a) and (c), horizontal dotted lines). Nevertheless,
for $s=0,$ i.e., for the counting probabilities $\{P_{n}(t)\}_{n=0}^{\infty
},$ one can also find a discretization time where the intensity develops an
intermittence phenomenon, i.e., there exist successive time intervals where
either many events happen or none happen at all.

In the thermodynamic picture, the active and inactive periods are associated
to two different phases, where finite-size effects lead to the transitions
between them \cite{Large}. In the present approach, the intermittence
phenomenon can be related to the WTD $w(t)$ [Eq. (\ref{WaitBiExpo})]. For
the chosen parameter values, which satisfy $p\ll 1$ and $\gamma _{<}/\gamma
_{>}\ll 1,$ $w(t)$ develops two different time scales, where the weight of
one of them is much higher than the other one. This feature is evident in
Fig. 4 $(s=0).$ Hence, the origin of the intermittence property follows
straightforwardly from the previous defined algorithm \cite{budini}. The
active and inactive time periods arise respectively from the short and long
time behaviors of the WTD. Consistently, the intensity of the active periods
fluctuates around $\gamma _{>}\approx \gamma $ (dotted line in Fig. 4b).

The intensity realizations show that when $s$ departs from zero, the
intermittence phenomenon is lost. In the thermodynamic picture this property
means that one of the two phases becomes dominant or stable. Specifically,
the active and inactive phases are dominant for $s<0$ and $s>0$ respectively
[see Fig. 3(a)]. Here, these conclusions can be related to the structure of
the stationary WTD $w_{q}^{\infty }(\tau ,s)$ [Eq. (\ref%
{waitingSEstacionaria})].

The free energy $\Theta (s)$ goes to higher negative values for increasing
negative values of $s$ [see Fig. 2]. Hence, for $s<0$ the two time scales
property is lost, and the exponential term $\exp [-\tau |\Theta (s)|]$
becomes dominant in Eq. (\ref{waitingSEstacionaria}). In consequence the $s$%
-ensemble, by using Eq. (\ref{LimitTheta}), can be approximated by a Poisson
process $w_{q}^{\infty }(\tau ,s)\approx \gamma \exp [-\gamma \tau ],$ i.e.,
the active phase. The Poisson property explains why for $s<0$ all
(normalized) response functions, independently of $s_{0},$ converge to the
same asymptotic curve [Fig. 3(a)], $\left\langle \left\langle N\right\rangle
\right\rangle /\gamma \approx \langle \langle \Delta N^{2}\rangle \rangle
/\gamma \approx \exp [-s].$

For increasing values of $s>0$ the free energy is positive, inducing a
slower decay of $w_{q}^{\infty }(\tau ,s).$ As can be seen in Fig. 4, the
survival probability maintains two different time scales, nevertheless their
relative weights are similar. Therefore, in this case the intermittence
phenomenon is also lost. Furthermore, as the weight of the larger time scale
increases with $s,$ the intensity is diminished, i.e., the inactive phase
becomes dominant for increasing $s.$

From the previous analysis, we conclude that the abrupt changes of $%
\left\langle \left\langle N\right\rangle \right\rangle $ and $\langle
\langle \Delta N^{2}\rangle \rangle $ around $s\approx 0$ reflect both the
loss of the intermittence property for $|s|>0,$ and the transition between
the active and inactive phases. In the thermodynamic approach, the behavior
around $s\approx 0$ is read and estimated from a thermodynamic-finite-size
analysis \cite{Large}. Those results can be straightforwardly applied in the
present context after decomposing the partition function, Eq. (\ref%
{ZNonLocalEvolution}), as $Z(t,s)=Z_{A}(t)+Z_{I}(t),$ where the evolution of
each contribution (Active, Inactive) can be written as 
\begin{eqnarray}
\frac{dZ_{A}(t,s)}{dt} &=&-\gamma (1-e^{-s})Z_{A}(t,s)  \notag \\
&&-\Gamma _{A}Z_{A}(t,s)+\Gamma _{I}Z_{I}(t,s),  \label{ZMarkoviana} \\
\frac{dZ_{I}(t,s)}{dt} &=&+\Gamma _{A}Z_{A}(t,s)-\Gamma _{I}Z_{I}(t,s). 
\notag
\end{eqnarray}%
Here, $\Gamma _{A}=\beta ,$ $\Gamma _{I}=\eta -\beta =\gamma _{>}\gamma
_{<}/\gamma ,$ and the initial conditions must to be $Z_{A}(0,s)=1,$ $%
Z_{I}(0,s)=0.$ This splitting allows to read the counting process $%
\{P_{n}(t)\}_{n=0}^{\infty }$ as a stochastic modulated Poissonian one [see
Eq. (\ref{ZPoisson})], whose rate $\gamma _{st}(t)$ at random times adopts
the values $\gamma _{A}=\gamma $ and $\gamma _{I}=0.$ The probabilities of
assuming each value are governed by a classical master equation with
transition rates $\Gamma _{A}$ and $\Gamma _{I}.$

In a slow modulation limit \cite{Large}, i.e., when $\Gamma _{A}/\gamma \ll
1 $ and $\Gamma _{I}/\gamma \ll 1,$ Eq. (\ref{ZMarkoviana}) leads to an
intermittence phenomenon. In fact, these conditions guarantee the existence
of periods of time with many consecutive events and periods where no events
happen at all. The previous inequalities can be satisfied by demanding $p\ll
1$ and $\gamma _{<}\ll \gamma _{>},$ such that $\gamma _{<}/\gamma
_{>}\approx \mathrm{O}(p).$ Hence, from Eq. (\ref{definitions}) it follows $%
\Gamma _{A}\approx p\gamma _{<},$ and $\Gamma _{I}\approx \gamma _{<}.$ The
parameters chosen in Fig. 2 satisfy these constraints. On the other hand,
from Ref. \cite{Large} we know that the transition rates control the
thermodynamic finite-size effects. In fact, the wide $\sigma _{p}$ of the
peak of the second response function $\langle \langle \Delta N^{2}\rangle
\rangle $ [inset of Fig. 3(a)] here can be estimated as $\sigma _{p}\approx
2(\Gamma _{A}+\Gamma _{I})/\gamma \approx 2p$ (see Eq. (66) in Ref. \cite%
{Large}). Therefore, in the limit $p\rightarrow 0,$ a first order transition
is approached, i.e., the first and second response functions converge to a
discontinuous and a delta Dirac function respectively. Due to the validity
of the shift closure condition, after a change of characteristic parameters,
Eq. (\ref{mapeoExpor}), the transition may be observed at an arbitrary value
of the parameter $s$ (see Fig. 3).

In general, when the WTD involves more terms [Eq. (\ref{variasRates})], it
is not possible to find an equivalent description such as the one given by
Eq. (\ref{ZMarkoviana}). Nevertheless, a similar approximated description
can be formulated when the intermittence phenomenon develops.

\section{Non-renewal counting processes}

While the renewal processes admit a relatively simple description, a general
formalism for dealing with non-renewal counting process does not exist. In
fact, the presence of memory between successive events does not admit a
general description. Nevertheless, many cases can be mapped with a situation
where the probability distribution for the interval between consecutive
events (WTD) changes randomly after each event \cite{vreeswijk,SMSJumps}.

We introduce a WTD $w(\tau ,x_{b}|x_{a})$ that, in contrast to the renewal
case, also depends on an extra unobservable (hidden) variable $\mathbf{x}$
whose states are $\{x_{i}\}.$ With $x_{a}$ and $x_{b}$ we denote the states
of the hidden variable before and after an event. Then, $w(\tau ,x_{b}|x_{a})
$ not only provides the statistics of the time intervals but also defines
the probability transition between the states of $\mathbf{x.}$ The most
simple case corresponds to $w(\tau ,x_{b}|x_{a})=w(\tau
,x_{a})T(x_{b}|x_{a}),$ where $w(\tau ,x_{a})$ is a WTD parametrized by $%
x_{a}$ and $T(x_{b}|x_{a})$ is the transition matrix between the hidden
states. Processes with memory of previous events, processes with adaptation
and double stochastic processes can be covered with this approach \cite%
{vreeswijk}. Furthermore, a stochastic WTD may arise when describing quantum
dissipative dynamics driven by classical fluctuations. In this last case $%
x_{a}$ and $x_{b}$ correspond to probabilities $\rho _{a}$ and $\rho _{b}$
of the initial and posterior hidden configurations \cite{SMSJumps}.

By introducing a vector space spanned by the states of $\mathbf{x,}$ $%
\{|x)\},$ and by using the definition of $w(\tau ,x_{b}|x_{a}),$ the
counting probabilities can be written as 
\begin{equation}
P_{n}(t)=(1|\hat{P}_{n}(t)|\rho _{in}).  \label{CountingVectorial}
\end{equation}%
Vector $|\rho _{in})$ denotes the initial probabilities of the hidden
variables, while $(1|\equiv (1,\cdots ,1).$ The operators $\hat{P}_{n}(t)$\
follow from a convolution%
\begin{equation}
\hat{P}_{n}(t)=\int_{0}^{t}\!dt_{n}\int_{0}^{t_{n}}\!dt_{n-1}\cdots
\int_{0}^{t_{2}}\!dt_{1}\hat{P}_{n}[t,\{t_{i}\}_{i=1}^{n}],
\label{CountingVect}
\end{equation}%
where the \textquotedblleft joint probability operator\textquotedblright\
reads%
\begin{equation}
\hat{P}_{n}[t,\{t_{i}\}_{i=1}^{n}]=\hat{P}_{0}(t-t_{n})\hat{W}%
(t_{n}-t_{n-1})\cdots \hat{W}(t_{1}).
\end{equation}%
The survival operator is $\hat{P}_{0}(t)=1-\int_{0}^{t}dt^{\prime }\hat{W}%
(t^{\prime }),$ and $\hat{W}(t)$ has components%
\begin{equation}
(x_{j}|\hat{W}(t)|x_{k})=w(t,x_{j}|x_{k}).
\end{equation}%
With the previous prescriptions, it is simple to derive similar expressions
for the $s$-ensemble [Eq. (\ref{q})]%
\begin{equation}
q_{n}(t,s)=(1|\hat{q}_{n}(t,s)|\rho _{in}),
\end{equation}%
where now the operators read%
\begin{equation*}
\hat{q}_{n}(t,s)=\int_{0}^{t}\!dt_{n}\int_{0}^{t_{n}}\!dt_{n-1}\cdots
\int_{0}^{t_{2}}\!dt_{1}\hat{q}_{n}[t,\{t_{i}\}_{i=1}^{n},s],
\end{equation*}%
with the joint-probability operator%
\begin{equation*}
\hat{q}_{n}[t,\{t_{i}\}_{i=1}^{n},s]=\hat{q}_{0}(t,t_{n},s)\hat{W}%
_{q}(t,t_{n-1},s)\cdots \hat{W}_{q}(t_{1},0,s).
\end{equation*}%
The survival operator reads $\hat{q}_{0}(t,t^{\prime },s)=\hat{P}%
_{0}(t-t^{\prime })Z(t^{\prime },s)/Z(t,s),$ while the waiting time operator
is 
\begin{equation}
\hat{W}_{q}(t,t^{\prime },s)=e^{-s}\hat{W}(t-t^{\prime })\frac{Z(t^{\prime
},s)}{Z(t,s)},  \label{WaiterVectoriales}
\end{equation}%
The partition function reads%
\begin{equation}
Z(t,s)=(1|\hat{Z}(t,s)|\rho _{in}),
\end{equation}%
where its associated operator is%
\begin{equation}
\hat{Z}(t,s)=\sum_{n=0}^{\infty }\hat{P}_{n}(t)e^{-sn}.
\end{equation}%
These expressions imply that the counting process defined by the
probabilities $\{q_{n}(t,s)\}_{n=1}^{\infty }$ has a similar structure to
that of the set $\{P_{n}(t)\}_{n=1}^{\infty }.$ In fact, the transformation
of the waiting time operator [Eq. (\ref{WaiterVectoriales})]\ is similar to
that of the renewal case [Eq. (\ref{qWaiting})]. Hence, in the long time
regime, it is also possible to define a stationary waiting time operator $%
\hat{W}_{q}^{\infty }(\tau ,s)=\lim_{t\rightarrow \infty }\hat{W}%
_{q}^{\infty }(t,t-\tau ,s),$ which reads%
\begin{equation}
\hat{W}_{q}^{\infty }(\tau ,s)=e^{-s}\hat{W}(\tau )\exp [\tau \Theta (s)].
\label{WaiterVectorial}
\end{equation}%
Thus, the non-renewal structure, in spite of the exponential decay measured
by $\Theta (s),$ remains the same. Similar results to that obtained for
renewal processes can be established from this last equation.

\section{Summary and conclusions}

The LD method allows to characterize the statistics of a point process with
a thermodynamic frame that is defined by the scaling rates of the counting
probabilities and its characteristic function. The basis of the present
analysis consists in finding an auxiliary process that maximizes the entropy
function of the thermodynamic approach, Eq. (\ref{q}). The conditional
measurement scheme defined in Fig. 1 provides its ensemble of stochastic
realizations from those of the original one. This result gives an
alternative measurement interpretation of the thermodynamic frame and its
conjugate counting field. Furthermore, it permits to relate the large
fluctuations of a counting process with the statistical properties of a
subset of its realizations.


Diverse general results were established under a renewal property, i.e.,
when the successive events do not develop any memory between them. We
established that the $s$-ensemble, independently of the structure of the WTD
of the counting process, is also a renewal process. An exponential decay
measured by the free energy function determines its (stationary) WTD [Eq. (%
\ref{waitingSEstacionaria})]. This function completely determines the
thermodynamic formalism. In fact, a general equation for the free energy can
be established in terms of its Laplace transform, Eq. (\ref{WaitTeta}).
Similarly, the response functions follow from its normalized moments, Eq. (%
\ref{ResponseRenewal}).

Different phenomena that appear in the thermodynamic formalism were analyzed
and derived in terms of the developed approach. We demonstrated that an
invariance scale property, Eq. (\ref{ScaleInvariant}), is satisfied whenever
the $s$-ensemble, in the long time regime, follows from a time rescaling of
the original one, Eq. (\ref{ShiftTime}). Equivalently, the invariance scale
property arises when the counting process can asymptotically be approximated
by a renewal one defined by a Gamma distribution, Eq. (\ref%
{WaiterInvariateEscalaLaplace}). A shift closure property was introduced
[Eq. (\ref{Closure})]. Its fulfillment implies that all possible statistical
behaviors of the $s$-ensemble can be achieved and generated from the
original one after an adequate redefinition of its characteristic
parameters. Alternatively, it means that there exists a transformation of
the characteristic parameters of the counting process whose sole effect is
to shift all thermodynamic response functions in the $s$-direction.
Different families of WTD guarantee the fulfillment of this property, Eqs.\ (%
\ref{fluoreto}) and (\ref{variasRates}). Finally, we showed that renewal
processes can in fact develop intermittent phenomena, where the realizations
of the process and the $s$-ensemble may be characterized by intervals of
high and null counting rate (Fig. 4). The properties of the response
functions, which approach a first order transition, were related to the
stochastic dynamics of the $s$-ensemble. While no general results can be
established for non-renewal processes, we showed that similar considerations
can be derived when the process is defined by a stochastic WTD, Eq. (\ref%
{WaiterVectorial}).

The present approach gives a solid basis for understanding the LD
thermodynamic frame of a counting process. Its main advantage is the
possibility of getting some general conclusions that apply to a wide class
of point processes, independently of its specific structure. Extension to
bidirectional counting processes, as well as non-ergodic ones may, in
principle, be done along similar lines.

\section*{Acknowledgments}

The author thanks fruitful discussions with E. Urdapilleta and to R.S.
Echeveste for a critical reading of this manuscript. This work was supported
by CONICET, Argentina, PIP 11420090100211.

\end{document}